\documentclass[11pt]{article}
\usepackage{amsfonts}
\usepackage{epsfig}
\newcommand{\be}{\begin{equation}}
\newcommand{\ee}{\end{equation}}
\newcommand{\bea}{\begin{eqnarray}}
\newcommand{\eea}{\end{eqnarray}}

\DeclareMathSymbol{\mg}{\mathrel}{symbols}{"1D}

\newcommand{\gl}{\lambda}

\newcommand{\cF}{{\cal F}}

\newcommand{\cL}{{\cal L}}
\newcommand{\cM}{{\cal M}}

\newcommand{\cO}{{\cal O}}

\newcommand{\cR}{{\cal R}}

\newcommand{\ra}{\rightarrow}

\newcounter{oldcounter}

\if@twoside\oddsidemargin25mm
\else
\begin{document}
\begin{flushright} 
DAMTP-2002-97
\end{flushright} 
\vskip 3cm
\begin{center} 
{\Large {\bf U(1) masses in intersecting  D-brane SM-like models}}
\\
\bigskip 
\vspace{0.93cm} 
{\large 
{\bf D.M. Ghilencea\footnote{
{{ {\ {\ {\ E-mail: D.M.Ghilencea@damtp.cam.ac.uk}}}}}}}}
\\
\vspace{0.9cm} 
{\it DAMTP, CMS, University of Cambridge} \\
{\it Wilberforce Road, Cambridge, CB3 0WA, United Kingdom.}\\
\bigskip 
\end{center}
\vspace{2cm}
\begin{center}
{\bf Abstract}\\
\vspace{0.5cm}
\end{center}
{\small{For recently constructed classes of D6-brane models, 
yielding the Standard Model fermion spectrum and gauge symmetry, 
we  compute lower bounds on the masses of  new $U(1)$ fields
that such models predict in addition to the hypercharge $U(1)_Y$.
In models with extra dimensions, generic uncertainties due to unknown 
values of the compactification radii  of the extra dimensions 
affect the value of the string scale and thus  the  
predictive power of such models. Using  $\rho$ parameter and 
$Z-U(1)$ mixing-angle constraints we show how to avoid such uncertainties, 
to provide  lower mass  bounds for  the additional $U(1)$ fields. These 
are in the region above 750 GeV for  mixing angles less than
 $1.5 \times 10^{-3}$ (and as low as 550 GeV for 
mixing angles of $3\times 10^{-3}$).}}

\newpage\setcounter{page}{1}

\section{Introduction.}
The interest in  physics of large extra dimensions triggered  an intense
research activity  at both effective field theory 
and string theory  level as well. Recent progress in the latter 
included  specific (intersecting) 
D-branes  string  constructions which were particularly 
successful in yielding in the low energy limit  a spectrum and 
symmetry close to that of the Standard Model (SM) \cite{imr}
(or supersymmetric extensions \cite{Cvetic:2002wh}). 
Such constructions were extensively analysed at string level,
and  may provide us with consistent, possibly realistic
models. However,  addressing the phenomenological implications
of such D-brane SM-like models is  at an early stage, 
thus motivating  this work as a step in this direction.

Chiral  D-brane models of possible phenomenological interest
correspond to D-branes at singularities \cite{Quevedo1,aiqu,cuw} and
D-branes intersecting at non-trivial angles 
\cite{imr,bgkl,afiru,afiru2,bkl,csu,bklo,pheno,cim1,cim2,Cremades:2002dh}.
Such models share some common properties:  a value  for the string 
scale which is lower than in the heterotic case,  
the  gauge symmetry includes direct products of groups  $U(N_\alpha)\times
U(N_\beta)$  (each $U(N_\alpha)$ emerging from the set/stack ``$\alpha$''
of $N$ individual $U(1)$ branes),  
the matter fields  transform as bi-fundamental representations of 
such products (as $(N_\alpha, {\overline N_\beta})$ or 
$(N_\alpha, N_\beta)$),  $U(1)$ factors originating each from 
initial $U(N)$ groups are a generic presence beyond the SM gauge group, 
and three generation models close to the 
Standard Model or supersymmetric extensions may be obtained.

In the present work  we consider  for our phenomenological investigation
the class of models of \cite{imr} with the purpose of extending 
the analysis started in \cite{Ghilencea:2002da}. 
These models are  non-supersymmetric string constructions
which  predict a fermion massless  spectrum {\it identical} to that
of the Standard Model. The Standard Model gauge group
emerges  from four stacks of D6-branes wrapping
a three cycle on a six-torus (orientifolded) type II A string
theory, in the presence of a background NS 
B-field. Similar constructions may be possible with D5 branes in type 
II B orientifold compactifications on an orbifold 
$T^2\times T^2\times T^2/Z_N$, and their analysis follows the 
pattern presented in this work 
\cite{Cremades:2002dh}\footnote{For other models  see 
also \cite{Antoniadis:2001np}.}.
The resulting gauge group  $U(3)\times U(2)\times U(1)\times U(1)$  
consequently contains four initial $U(1)$ groups in addition to the
non-Abelian part of the Standard Model gauge group. These may
be identified with the  baryon number, lepton number, Peccei-Quinn-like
symmetry and hypercharge (or linear combinations thereof).

Upon dimensional reduction to four dimensions,
three  (linear combinations) of these additional $U(1)$ fields
become massive through  couplings $B_i\wedge F$ where $B_i$ 
stands for four dimensional  RR two-form fields (present in the models 
of \cite{imr}) and F for the $U(1)$ field strength
tensor, as detailed in \cite{Ghilencea:2002da}. This mechanism 
applies to anomalous $U(1)$'s as well\footnote{The anomalies are cancelled by a
generalised Green-Schwarz mechanism, see ref. \cite{imr}.}
as to $U(1)$ fields which are {\it not} anomalous, and does not 
involve a Higgs mechanism/particle. Consequently, after $U(1)$ fields become
massive, the corresponding $U(1)$ symmetries remain as  perturbatively exact 
{\it global} symmetries in the effective Lagrangian.  For example baryon number
remains  a  global symmetry which  may be welcome to 
explaining  the stability of the proton (troublesome in
models with a low string scale).  The need for a low string scale
is even more important for  the class of models 
considered is non-supersymmetric, requiring a value for  $M_S$
in the region of few TeV,  to avoid a hierarchy problem. 
The situation is unlike that of heterotic counterpart (and
supersymmetric) models, where  baryon number violation interactions
may instead be suppressed by the existence of a high UV (string) scale.

The class of models of \cite{imr} ``accommodate'' three generations of matter
fields and this  is briefly justified in the following way.
The tadpole cancellation condition requires the number of fundamental 
$N_a$ and anti-fundamental $\overline N_a$  representations for any
$U(N_a)$ group be equal, even if the gauge group is $U(2)$.
This restricts the assignment of quarks and leptons as  $U(2)$ doublets or
anti-doublets. Indeed, if all left-handed quarks were $U(2)$ doublets, 
and leptons were $U(2)$ anti-doublets, one could not satisfy in this case
$\#_{N_a}=\#_{{\overline N_a}}$. The  only possibility is to have  two 
left-handed quarks $Q_L^i$ i=1,2 as $U(2)$ doublets (anti-doublets) 
with  the third one and with left-handed leptons as anti-doublets (doublets). 
For {\it three} families of quarks/leptons the total number of doublets and 
anti-doublets is in this way equal, ensuring tadpole
cancellation and relating  the  number of colours to that of
generations \cite{imr}.

The class of models we address  being non-supersymmetric require
 a low ($TeV$ region) string scale to avoid a hierarchy problem.
In \cite{Ghilencea:2002da} it was  shown
that such a low value of the string scale 
($M_S\approx 1.5 \,TeV$) can indeed be obtained while still 
complying with precision electroweak measurements.  
It remains to explain why the string scale may take
such low values. The usual explanation is that some transverse
dimensions become very large. This can apply to D5 (toroidal-like)
 models \cite{Cremades:2002dh}, but  in the case of models with 
D6 branes discussed in \cite{imr}   there is no compact dimension which
is simultaneously transverse to all the SM branes. Possible 
solutions were suggested in  \cite{Ghilencea:2002da} and in fact the 
problem is elegantly solved in models with D6 branes wrapped on 
3-cycles\footnote{localised in a small region of the Calabi-Yau manifold.}
in the more general Calabi-Yau compactifications \cite{Uranga1}.

One potential drawback of the models (both D5 and D6 branes) we 
refer to may be that the models cannot comply with the successful
unification of the gauge couplings of the supersymmetric version of 
the SM (MSSM). This is however expected since the models  have
SM-like spectrum and are  also non-supersymmetric. 
The problem is actually more general and present even in
supersymmetric models,  for cases with a low (TeV range) string 
scale/large extra dimension(s) \cite{Ghilencea:2000dg}.

In the next section we review the mechanism by which the $U(1)$ fields 
become massive and describe the procedure we use for setting explicit lower
bounds on the masses of $U(1)$ bosons. Finding such lower bounds on the
$U(1)$ masses is the main purpose of this work. The results are
presented in  Section \ref{lowerbounds}. 
This completes the analysis of \cite{Ghilencea:2002da} which 
only addressed the  bounds on the value of the string scale complying 
with constraints from electroweak scale measurements ($\rho$ parameter).
Finally, the  lower  bounds on $U(1)$ masses are  compared to those 
from  other (string) models predicting additional Z' bosons 
in the low energy regime.

\section{$U(1)$ masses from string theory.}

The class of D6 models that we investigate in this work
 are constructed  from Type II A string theory compactified on 
a six-torus $T^2\times T^2 \times T^2$.  Apart from  the Minkowski
space, the remaining three dimensions of $D6$ brane models are wrapped each
on  a different torus  $T^2$. One actually considers four stacks of branes, 
$D6_{\alpha}$ branes, $\alpha=a,b,c,d$, each stack $\alpha$ bringing a
$U(N_\alpha)$ group. Further, $n_{\alpha i} \, (m_{\alpha i})$
 $i=1,2,3$ stand for  the  wrapping numbers 
of each $D6_{\alpha}$ around the x (y) coordinate of the i-th two-torus. 
The general set of allowed wrapping numbers yielding the Standard
Model spectrum was presented in \cite{imr} with remaining
independent parameters given in Table \ref{minimal}. 
In addition to the non-Abelian part of the SM gauge group, 
a $U(1)_\alpha$ factor emerges from  each stack of branes. Therefore 
the final gauge group is 
$SU(3)\times SU(2)\times U(1)_a \times U(1)_b\times U(1)_c \times
U(1)_d$ with the hypercharge group to arise as a linear combination
of the  four initial $U(1)_\alpha$.
The charges of SM fields under $U(1)_\alpha$  are given
in the Appendix, Table \ref{tabpssm}. For a full description of the 
models see \cite{imr} with applications in  \cite{Ghilencea:2002da}.

Although we do not  investigate them in the following, similar 
constructions exist in models with D5 branes as well.
D5-brane orientifold models are obtained from Type II B
compactifications on an orbifold $T^2\times T^2 \times T^2/Z_N$.
Four stacks (a,b,c,d) of D5 branes are wrapped on cycles of $T^2\times T^2$ and
are located at a $Z_N$ fixed point of $T^2/Z_N$. The six dimensional
world volume includes the Minkowski space with the remaining two extra
dimensions  wrapping  each a different two-torus. Each stack of branes
is then specified by $n_{\alpha i} \, (m_{\alpha i})$ i=1,2. One obtains the
SM fermionic spectrum and a gauge group with four additional $U(1)_\alpha$
factors, similar to D6 brane models. The gauge group is again
$SU(3)\times SU(2)\times U(1)_a \times U(1)_b\times U(1)_c \times
U(1)_d$ with hypercharge to emerge as a linear combination of 
the four Abelian groups. Details of D5 models may be found in
\cite{Cremades:2002dh}.

For both D6- and D5-brane models there are four RR two-form fields $B_i$ 
which couple to the field strength tensor of the four 
$U(1)_\alpha$ fields, in the following way
\begin{equation}\label{action}
\cL\supset \sum_{i=1}^{4} \sum_{\alpha} 
c_i^\alpha B_i\wedge Tr F^{\alpha}, \qquad
\quad \alpha=a,b,c,d. 
\end{equation}
Such couplings can provide masses to three linear
combinations of $U(1)$ fields but not to the (linear combination
yielding the) hypercharge $U(1)_Y$ field,  which  remains  a local
symmetry of the theory, before electroweak symmetry breaking.  
As a consequence  not all ($4\times 4=16$) 
coefficients $c_i^\alpha$ are independent. It turns out that
only seven of them are non-vanishing and five of them are actually
independent. Their explicit expressions were presented in 
\cite{Ghilencea:2002da} for the D6 and D5 brane models considered
and are also included in the Appendix.

The exact mechanism by which $U(1)_\alpha$ fields become massive, 
enabled by the term  (\ref{action}) and by kinetic terms for $B_i$
was  presented in \cite{Ghilencea:2002da}.
The idea is that the kinetic term of $B_{i, \mu\nu}$ in the action 
combines with the  couplings (\ref{action}) to give 
in the action a mass term for the $U(1)$ fields and a corresponding gauge
kinetic term. This is just a re-arrangement of the degrees of freedom 
of 4D two-form fields $B_i$ whose scalar dual is each ``eaten'' by 
a $U(1)_\alpha$ field. 
Thus the mechanism does not require introducing  additional scalar
fields with vacuum expectation values, to provide mass
terms for the  $U(1)$ gauge bosons, nor does it bring the presence 
of massive Higgs-like fields in the end. Thus the  mechanism is 
different from the usual Higgs mechanism.
The following mass terms for $U(1)_\alpha$ fields emerge
\begin{equation}\label{masones}
\cL\supset 
\frac{1}{2}\sum_{\alpha,\beta} (M^2)_{\alpha\beta} A_\alpha A_\beta\equiv
  \frac{1}{2}\sum_{\alpha,\beta} 
\bigg[g_\alpha g_\beta M_S^2 \sum_{i=1}^{3} c_i^\alpha
c_i^\beta\bigg] \, A_\alpha A_\beta,\,\,\,\,\,\,\,
\alpha,\beta=a,b,c,d.
\end{equation}
The sum over i runs over the  three (massive) RR-fields
present\footnote{The hypercharge $U(1)_Y$ remains massless since there is no  
coupling $B_i\wedge F_y$ where $F_y$ is a linear combination of the
initial four $U(1)_\alpha$ fields strengths. 
Due to this the sum in (\ref{masones}) runs only over i=1,2,3, unlike in
(\ref{action}). This is possible because  
at string level there always exists one {\it massless}
$U(1)$ which together with the model building 
constraint  in the second-last column of Table
\ref{minimal} may be identified with $U(1)_Y$. The origin of the
existence of a massless (gauged) $U(1)$ at string level 
is not clear, but it can be related to topological arguments.}
in the models and $g_\alpha$ is the coupling of $U(1)_\alpha$.
($U(1)_a$ arises from $U(3)$ thus $g_a^2=g_{QCD}^2/6$ and 
$U(1)_b$ arises from $U(2)$ thus $g_b^2=g_L^2/4$).
Upon diagonalisation the mass matrix leads to positive 
(masses)$^2$, $M_i^2>0$ (i=2,3,4) ($M_1^2=0$ for hypercharge) 
for the three  $U(1)$  bosons.

The coefficients $c_i^\alpha$ in (\ref{masones}) depend on 
the normalisation of the kinetic terms for the RR fields $B_i$.
Such kinetic terms are  in general radii dependent and
once one redefines the fields to canonical kinetic terms, extra
volume factors $\underline\xi^i$ appear in (\ref{masones}) multiplying each
$c_i^\alpha$. It is  difficult to estimate the (numerical) value of such factors
on string theory grounds, and this is reducing the
predictive power of the models considered.
Their expressions in terms of the radii $R_{1,2}^i$ of torus $i$ are
for  the D6-brane case\footnote{Similar relations apply for D5-brane case.}:
\begin{equation}\label{xxii}
\underline\xi^i=\bigg[\frac{R_2^i R_1^j R_1^k}{R_1^i R_2^j R_2^k}\bigg]^{1/2},
\qquad i\not=j\not=j\not=k\not=i
\end{equation}
As in  \cite{Ghilencea:2002da} we assume that the factors 
$\underline\xi^i$ are equal in magnitude and thus may be absorbed
into the  re-definition of the string scale $M_S$ in
eq.(\ref{masones}). As a consequence
the string scale prediction is subject to the uncertainty
induced by such volume factors  \cite{Ghilencea:2002da}. 
For the remaining of this work $M_S$ will thus stand for this 
{\it re-scaled} value.

The natural question that emerges is then whether one is indeed able to
avoid the constraints induced by the (unknown) volume factors 
$\underline\xi^i$  and make a {\it prediction} in the class of models
under discussion. A low value of the string scale 
as found in \cite{Ghilencea:2002da} shows that one can construct 
string models with $M_S$ in the region of few TeV, while still
complying with current constraints from $\rho$ parameter physics. 
In itself this finding is important and reassuring  for the
consistency of the model, since it does not require a
string scale too large compared to TeV scale (which 
would re-introduce a hierarchy problem).
However, for experimental searches it is less important the 
exact value of $M_S$ which still complies with current experimental 
constraints, or the aforementioned uncertainties affecting it,
and more relevant the actual lower bounds on the  masses of the additional 
$U(1)$ fields.  It turns out that one 
can make a prediction for the latter independently of the volume factors
mentioned above, by only assuming that they are equal for
the three torii ($\underline\xi^1=\underline\xi^2=\underline\xi^3$) 
for D6 case (or two torii in case of D5 branes). This can be  respected  
if for example  the ratio of the two radii of the $i$-th torus,  
$R_2^i/R_1^i$ is the same for any $i=1,2,3$.

Using $\rho$ parameter constraints one can extract (lower)
bounds on the value of $M_S$, as already done in  \cite{Ghilencea:2002da} 
for D6 models.  Further, using the mass eigenvalues 
$M_{i}^2$ of $(M^2)_{\alpha\beta}$ computed in terms of $M_S$ 
(to which we add  their electroweak corrections) we are  able to predict 
lower bounds on the total masses $\cM_i$ of the additional $U(1)$ fields, 
{\it independent} of the volume factors $\underline\xi^i$. 
Therefore, while the string scale prediction is affected by such
unknown factors, one can make predictions for the
lower bounds on the masses of  $U(1)$ fields. This motivated the 
present analysis and  completes the  discussion of 
\cite{Ghilencea:2002da} which overlooked this observation.

\subsection{Lower bounds on $U(1)$ masses.}\label{lowerbounds}
For   D6 models the parameters are presented in Table \ref{minimal}
while the Higgs sector quantum numbers are presented in Table \ref{thirdone}.
After diagonalisation of the mass matrix (\ref{masones}) one computes 
the masses $M_i^2$ (i=2,3,4) ($M_1^2=0$ for hypercharge) and eigenvectors 
for gauge bosons,  according to the relations: 
\begin{equation}
A'_i = \sum_{a,b,c,d}\cF_{i\alpha} A_\alpha; \qquad
\delta_{ij} M_i^2 = \cF M^2 \cF^T, \qquad i,j=\overline{1,4}.
\end{equation}
where explicit entries for $\cF_{i\alpha}$ are given in the Appendix
eqs.(\ref{eigvecA}), (\ref{eigvecB}) and  $\cF\cF^T=\cF^T\cF=1$.

Upon electroweak symmetry breaking, the mass matrix (\ref{masones}) 
receives corrections from  the mixing of the Higgs state with
some  $U(1)_\alpha$ fields, present if  Higgs state is charged under 
$U(1)_\alpha$.  The fields $A_i'$ receive 
an additional  electroweak mass correction and $Z$ boson 
acquires a mass as well, a fraction  of which is due to 
the string mechanism for mass, induced by the mixing of Z 
boson with massive $U(1)$'s.
\begin{table}[t] 
\begin{center}
\begin{tabular}{|c|c|c|c|c|c|c|c|c|}
\hline
 Higgs    &  $\nu
 $  &  $\beta_1$   & $\beta_2$ & $n_{a2}$ & $n_{b1}$
& $n_{c1} $
& $n_{d2}$  & $N_h$ \\
\hline\hline  $n_H=1,n_h=0$  & 1/3  &  1/2  & $\beta_2$ & $n_{a2}$ & -1 & 1&
$\frac{1}{\beta_2}-n_{a2}$   &  $4\beta_2(1-n_{a2})$  \\
\hline  $n_H=1,n_h=0$  & 1/3  &  1/2  & $\beta_2$ & $n_{a2}$ & 1 & -1&
$-\frac{1}{\beta_2}-n_{a2}$   &  $4\beta_2(1-n_{a2})$  \\
\hline  $n_H=0,n_h=1$  & 1/3  &  1/2  & $\beta_2$ & $n_{a2}$ & 1 & 1&
$\frac{1}{\beta_2}-n_{a2}$   &  $4\beta_2(1-n_{a2})-1$  \\
\hline  $n_H=0,n_h=1$  & 1/3  &  1/2  & $\beta_2$ & $n_{a2}$ & -1 & -1&
$-\frac{1}{\beta_2}-n_{a2}$   &  $4\beta_2(1-n_{a2})+1$  \\
\hline\hline  $n_H=1,n_h=1$  & 1   &  1   & $\beta_2$ & $n_{a2}$ & 0  & 1&
$\frac{1}{3}(\frac{2}{\beta_2}-n_{a2})$   &
$\beta_2(8-\frac{4n_{a2}}{3})-\frac{1}{3}$ \\
\hline  $n_H=1,n_h=1$  & 1   &  1   & $\beta_2$ & $n_{a2}$ & 0  & -1&
$\frac{1}{3}(-\frac{2}{\beta_2}-n_{a2})$   &
$\beta_2(8-\frac{4n_{a2}}{3})+\frac{1}{3}$ \\
\hline  $n_H=1,n_h=1$  & 1/3   &  1   & $\beta_2$ & $n_{a2}$ & 0  & 1&
$ \frac{2}{\beta_2}-n_{a2}$   &
$\beta_2(8-{4n_{a2}})-1$ \\
\hline  $n_H=1,n_h=1$  & 1/3   &  1   & $\beta_2$ & $n_{a2}$ & 0  & -1&
$ -\frac{2}{\beta_2}-n_{a2}$   &
$\beta_2 (8-{4n_{a2}})+1$ \\
\hline \end{tabular}
\end{center}
\caption{\small Families of D6-brane models  
with minimal Higgs content as derived in \cite{imr}.
The first four lines correspond to models of
Class A of ref.\cite{Ghilencea:2002da}, the remaining ones to models
of Class B, which are distinguished by their different Higgs sector. 
The parameters of the models are  $n_{a2}$ and
$g_d/g_c$. The definition of $n_{d2}$ enables one to obtain the SM
value for hypercharges, and thus to identify the
massless (linear combination of) $U(1)$ with $U(1)_Y$.
$\beta_i=1\,(1/2)$ is a parameter due (in a  T-dual picture) to background 
NS B field (which modifies the complex structure of the
i$^{th}$ two-torus) and  corresponds to orthogonal
(tilted) torus. The presence of $\beta_i$ enables an odd number of
generations in these models  \cite{bkl} and also a shift in the
effective wrapping numbers. Finally $N_h$ stands for the number of 
branes parallel to the orientifold plane, added for global RR tadpole 
cancellation \cite{imr}.}
\label{minimal}
\end{table}
The new mass matrix $\cM^2_{\gamma\gamma'}$ in the ``extended'' basis
$a,b,c,d$ {\it and} $W_3^\mu$ (of $SU(2)_L$) contains  a $4\times 4$ 
sub-block  $M_{\alpha\beta}^2$ ($\alpha,\beta=a,b,c,d$)
to which we added electroweak  corrections. Its 
explicit form was presented in \cite{Ghilencea:2002da}. The
eigenvectors of $\cM^2_{\gamma\gamma'}$ 
(denoted by $\cF^*$) are computed in the Appendix eqs.(\ref{EWSB}) 
and satisfy
\begin{equation}
A_i^*  = \sum_{\gamma=a,b,c,d,W_3} {\cF^*}_{i\gamma} A_\gamma;\qquad
\delta_{ij} \cM_i^{2}  =  {\cF^*} \cM^2 {\cF}^{* T}, \qquad
i,j={\overline {1,5}};
\end{equation}
with $A_1^*$ to stand for the photon, $A^*_{2,3,4}$ for the massive
$U(1)$ fields, $A_5^*$ for   $Z$ boson 
and  $\cF^*\cF^{* T}=\cF^{* T}\cF^*=1$.
The mass of Z boson including the string corrections induced by
the term (\ref{masones}) is
\begin{equation}
M_Z^2\equiv \cM_5^2=
M_0^2\left[1+\eta \xi_{21}+\eta^2\xi_{31} + \cO(\eta^3)\right],\quad
M_0^2=\frac{1}{4}(4 g_b^2+g_y^2)<\!\phi\!>^2,\quad \eta=<\!\phi\!>^2/M_S^2
\end{equation}
where $M_0$ is the   mass of Z boson as given in the Standard Model, 
$<\!\phi\!>^2 \cos^2\theta \equiv H_1^2+H_2^2$,
$<\!\phi\!>^2 \sin^2\theta \equiv h_1^2+h_2^2$  
and  $\theta$ is the mixing angle in the Higgs sector.
$\theta=0,(\pi/2)$ for Class A models with 
$n_H=1,n_h=0$ ($n_H=0,n_h=1$) respectively.
For Class B  models $\theta$  is not restricted,
playing a role similar to $\tan\beta$ of the MSSM.
Imposing $\rho$ parameter constraints on the mass of $Z$ boson, 
we  derived the lower bound eq.(\ref{msnew}) on the
value of $M_S^2$ \cite{Ghilencea:2002da}. Note that $M_S^2$ 
includes possible volume factors effects ${\underline \xi}^i$ 
(see eq.(\ref{xxii}) and text thereafter).
\begin{equation}\label{msnew}
M_S^2= <\phi>^2 (- \xi_{21}) \left[1+\frac{\rho^0}{\Delta\rho}\right]
\left[1-\, 2\, \frac{\xi_{31}}{\xi_{21}^2}
\frac{\Delta\rho/\rho^0}{1+\Delta\rho/\rho^0}\right]^{1/2}
\end{equation}
The last bracket  brings a correction
to $M_S^2$ less than $0.1\%$ relative to the case of a full numerical
approach to computing $M_S^2$,  and will thus be ignored hereafter.
The value of $\xi_{21}$  is given by
\begin{eqnarray}
\xi_{21} & = & -\left\{
\beta_1^2 \left[2 \beta_1 g_y^2 \,\nu\, n_{c1} (1+\cR^2)-(36 g_a^2+g_y^2
\cR^2)\beta_2 n_{a2} \nu\right]^2+
4 \beta_1^2 \beta_2^4 \epsilon^2 (36 g_a^2+g_y^2 \cR^2)^2\right.\nonumber\\
&&\!\!\!\!\!\!\!\!\!\!\!\!\!\!\!\!\!
+\left. 9\beta_2^4\left[ g_y^2 n_{b1} \,\nu\, \cR^2 +12 g_a^2 
[ 3 n_{b1}\,\nu\,-n_{c1}(1+\cR^2)\cos(2\theta)]\right]^2
\right\} 
\left[5184\beta_1^2 \beta_2^2 \epsilon^2 g_a^4 n_{c1}^2
(1+\cR^2)^2\right]^{-1}
\end{eqnarray}
The above two equations give the lower bounds on $M_S^2$
in terms of the chosen  parameters, which are the ratio
$\cR=g_d/g_c$ and $n_{a2}$ while $g_y$ is the hypercharge 
coupling, eq.(\ref{hypercharge1}). 

\begin{table}[t] 
\begin{center}
\begin{tabular}{|c|c|c|c|c|c|c|c|c|c|}
\hline
Higgs $\sigma$  &  $q_b$   &  $q_c$   & $q_1'\equiv q_y$ & $T_3=1/2\,
\sigma_z$ &
Higgs $\sigma$  &  $q_b$   &  $q_c$   & $q_1'\equiv q_y$ & $T_3=1/2\, \sigma_z$ \\
\hline\hline
$h_1$   &    1     & -1       & 1/2              &   +1/2      & 
$H_1$   &   -1     & -1       & 1/2              &   +1/2    \\

\hline 
$h_2$   &   -1     &  1       &  -1/2            &   -1/2     & 
$H_2$   &    1     &  1       &  -1/2            &   -1/2     \\
\hline 
\end{tabular}
\end{center}
 \caption{Higgs fields, their U(1)$_{b,c}$
charges and weak isospin with $\sigma_z$  the diagonal Pauli matrix.
Class A models contain either $H_{1,2}$ or $h_{1,2}$ while Class B
contains both $H_{i}$ and $h_i$ with mixing angle $\theta$.}
\label{thirdone}
\end{table}

One constraint in the case of additional $U(1)$ 
bosons is that on their mixing angle with the usual $Z$ boson, which may be
read from the eigenvector of the latter.
For the case when only one additional $Z'$ 
boson exists, the mixing is induced by the presence of an off-diagonal mass
term $m_{ZZ'}^2 Z Z'$ in the action (in addition to $m'^2 Z'^2$ and
$m^2 Z^2$). This mixing 
may be expressed in function of the  mass eigenstates as 
(see for example \cite{Erler:1999nx})
\begin{equation}\label{oneboson}
\psi_0=\arctan\left[\frac{M_0^2-M_Z^2}{M_{Z'}^2-M_0^2}\right]^{1/2}
\end{equation}
with $M_{Z'}$ the mass of the additional boson.
Current experimental constraints provide bounds on the mixing angle
$\theta$ in the region of $\psi_0= k\times 10^{-3}$ with $k$ of
order unity \cite{Erler:1999nx}.

For more than one  additional boson which is our case, relation
(\ref{oneboson})  does not hold.  
In this case the mixing 
of $Z$ boson with the massive $U(1)$  fields can be computed 
using $\cF$ and $\cF^*$ of eqs.(\ref{eigvecA}), (\ref{eigvecB}), (\ref{EWSB}) 
to give
\begin{equation}\label{zz'} 
Z\equiv A_5^*=\sum_{i=1}^{4}\psi_i A_i' +\cF^*_{5 W_3}\, W_3,\qquad
\psi_i\equiv \sum_{\alpha=a,b,c,d}\cF_{5\alpha}^*\, \cF_{i\alpha}
\end{equation}
where  $\psi_i$ accounts
for the mixing $Z-A_i'$ in the  basis (normalised) 
of the  fields $A_i', W_3$, $i=1,2,3,4$. In the
limit of an infinite string scale, the massive $U(1)$ fields decouple
($\psi_{2,3,4}\ra 0$) to leave the usual mixing of the Standard Model 
with $\cF^*_{5W_3}\ra \cos\theta_W$ and $\psi_1\ra \sin\theta_W$. 
Eq.(\ref{zz'}) thus provides the eigenvector
of the Z boson for D6-brane SM-like models.
\begin{figure}[p] 
\begin{center}
\begin{tabular}{cc|cc|} 
\parbox{6.6cm}{ 
\psfig{figure=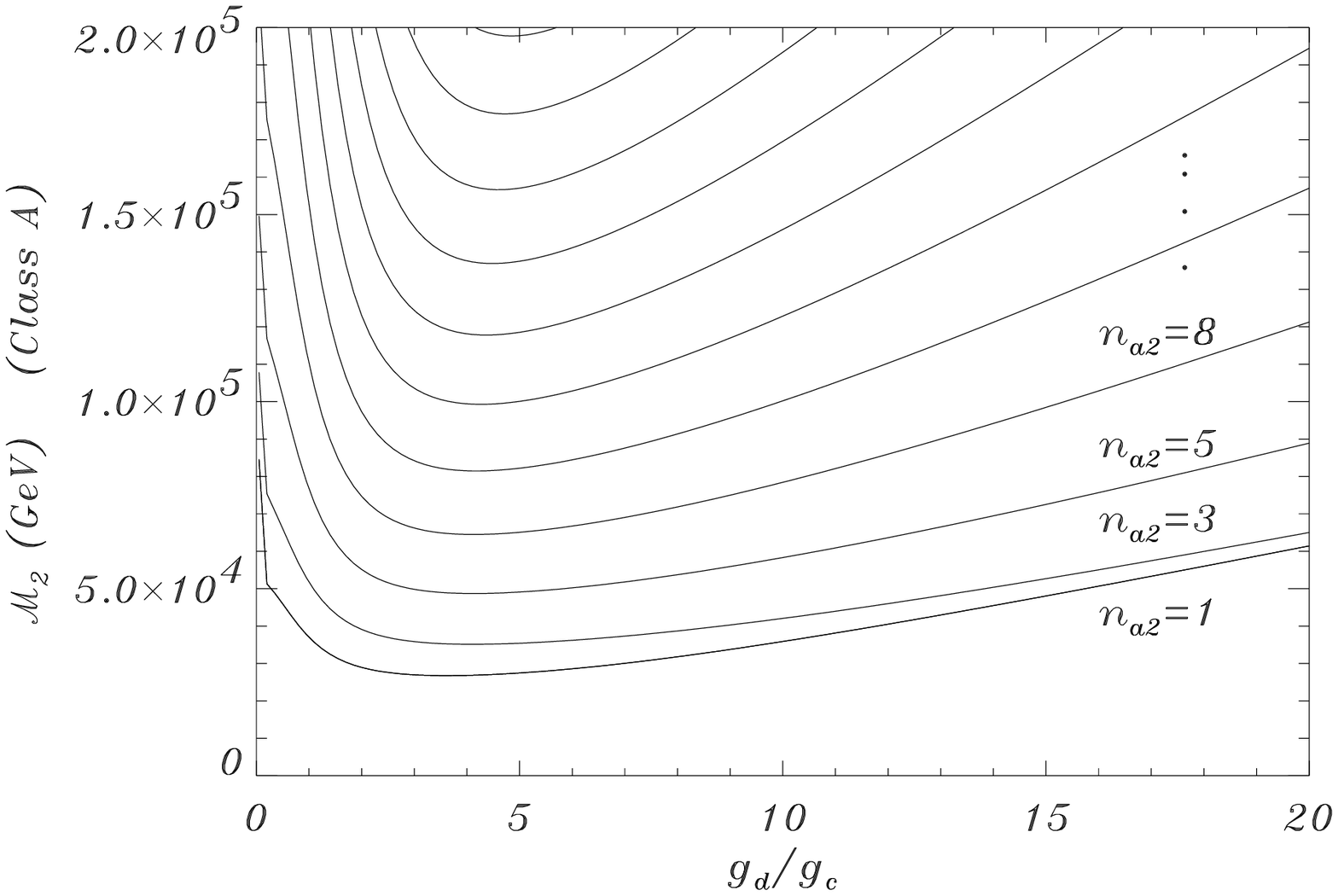,height=5.4cm,width=5.66cm}} 
\hfill{\,\,} 
\parbox{6.6cm}{ 
\psfig{figure=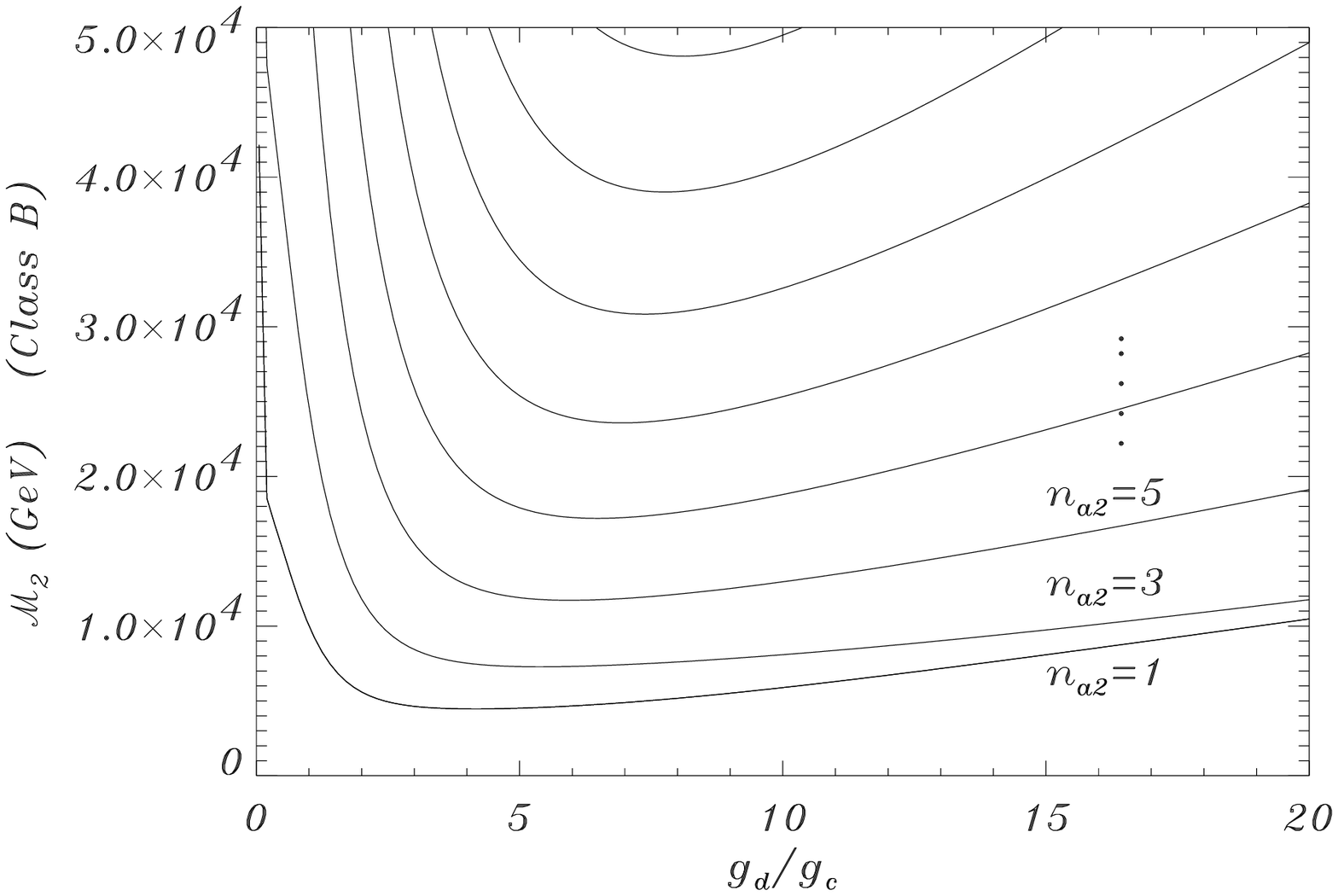,height=5.4cm,width=5.66cm}}  
\end{tabular} 
\begin{tabular}{cc|cc|} 
\parbox{6.6cm}{ 
\psfig{figure=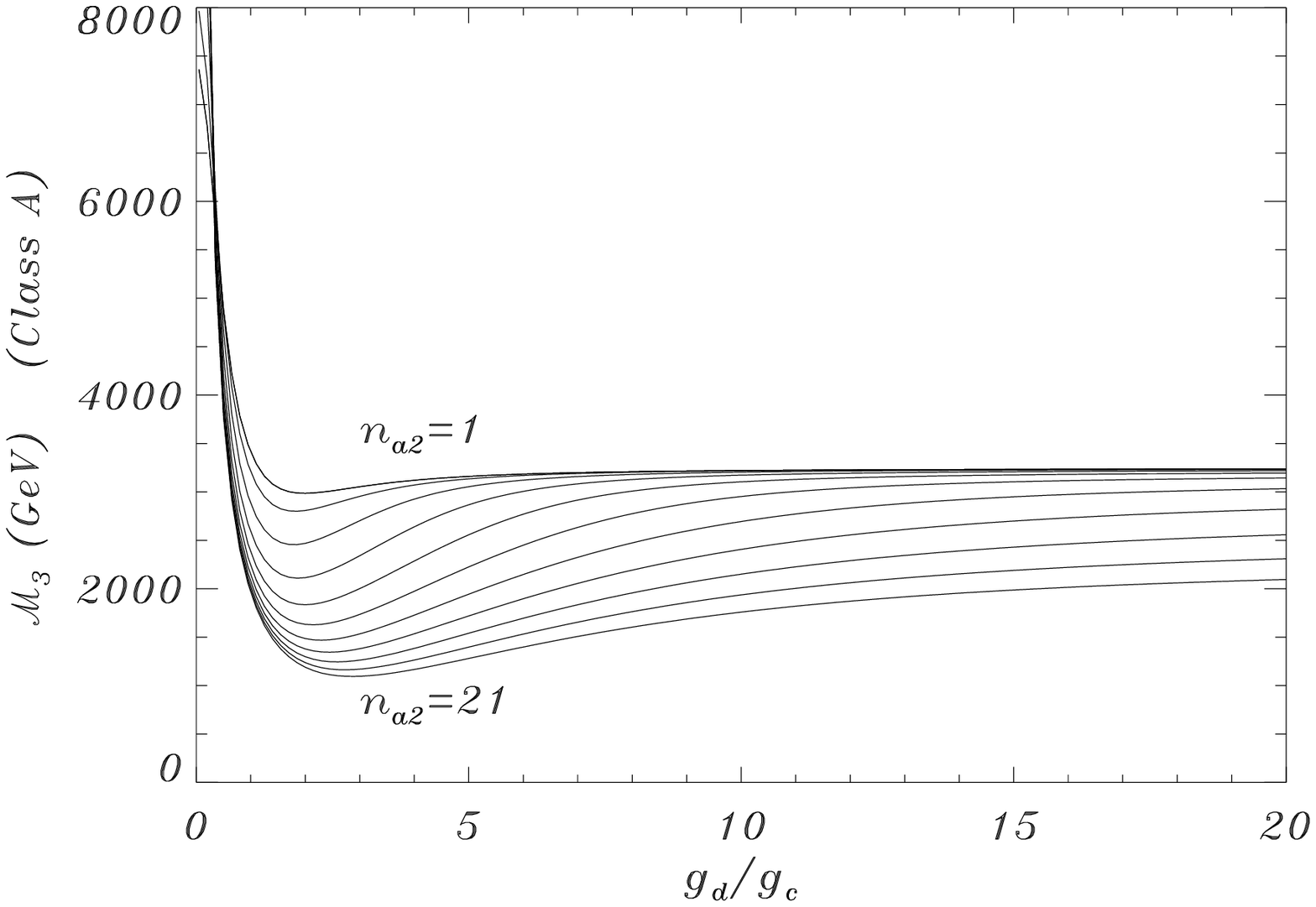,height=5.4cm,width=5.66cm}} 
\hfill{\,\,} 
\parbox{6.6cm}{ 
\psfig{figure=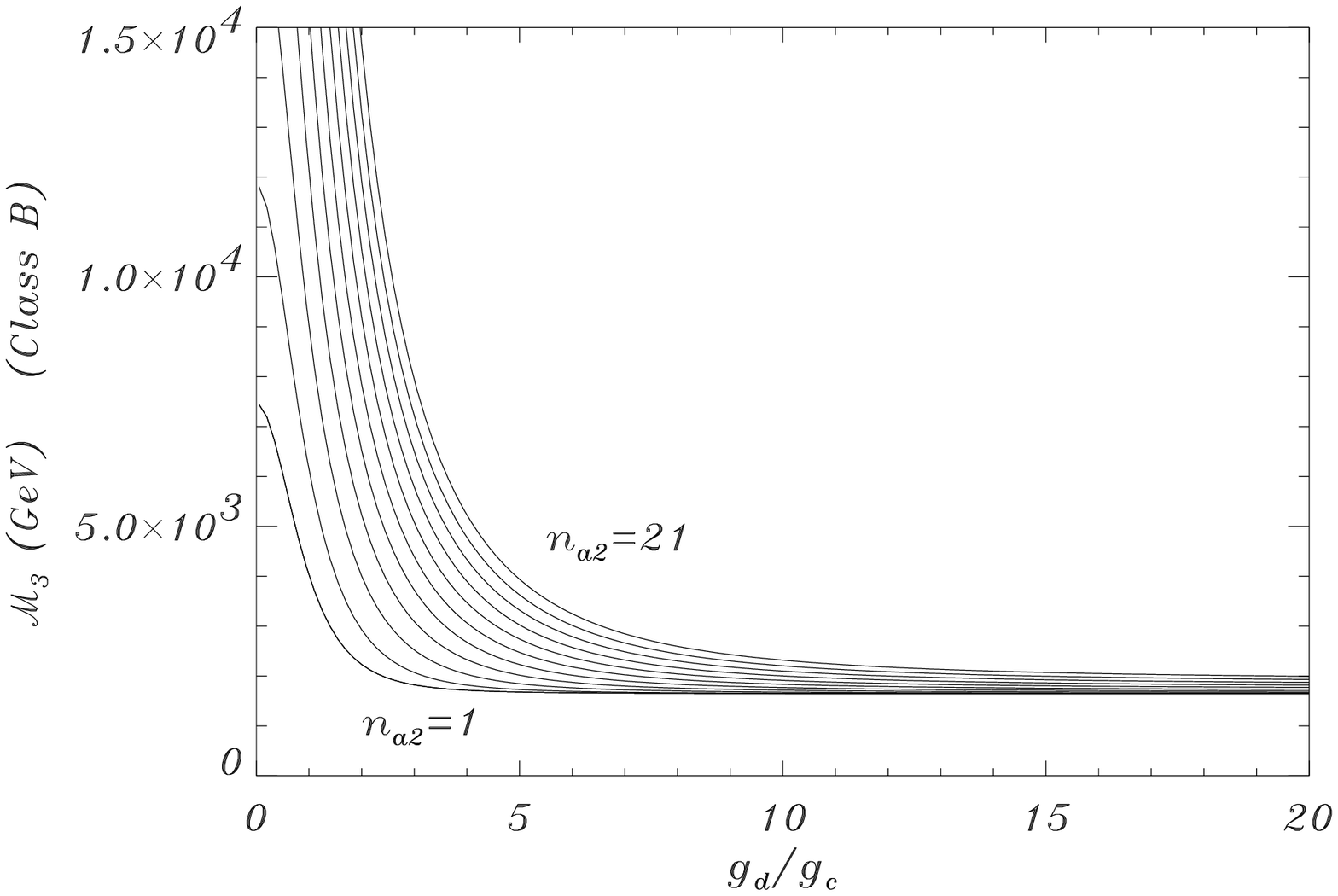,height=5.4cm,width=5.66cm}}  
\end{tabular} 
\begin{tabular}{cc|cc|} 
\parbox{6.6cm}{ 
\psfig{figure=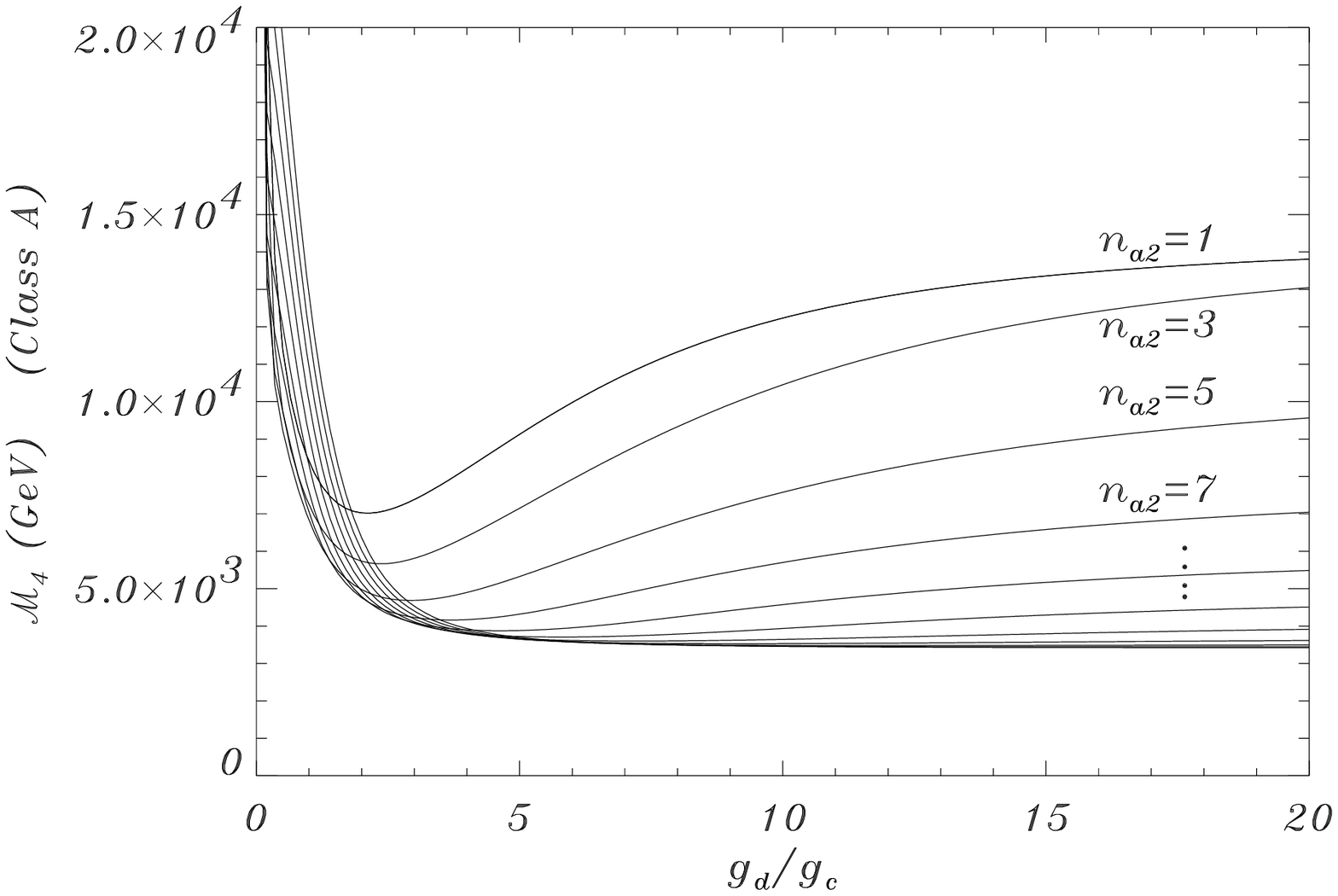,height=5.4cm,width=5.66cm}} 
\hfill{\,\,} 
\parbox{6.6cm}{ 
\psfig{figure=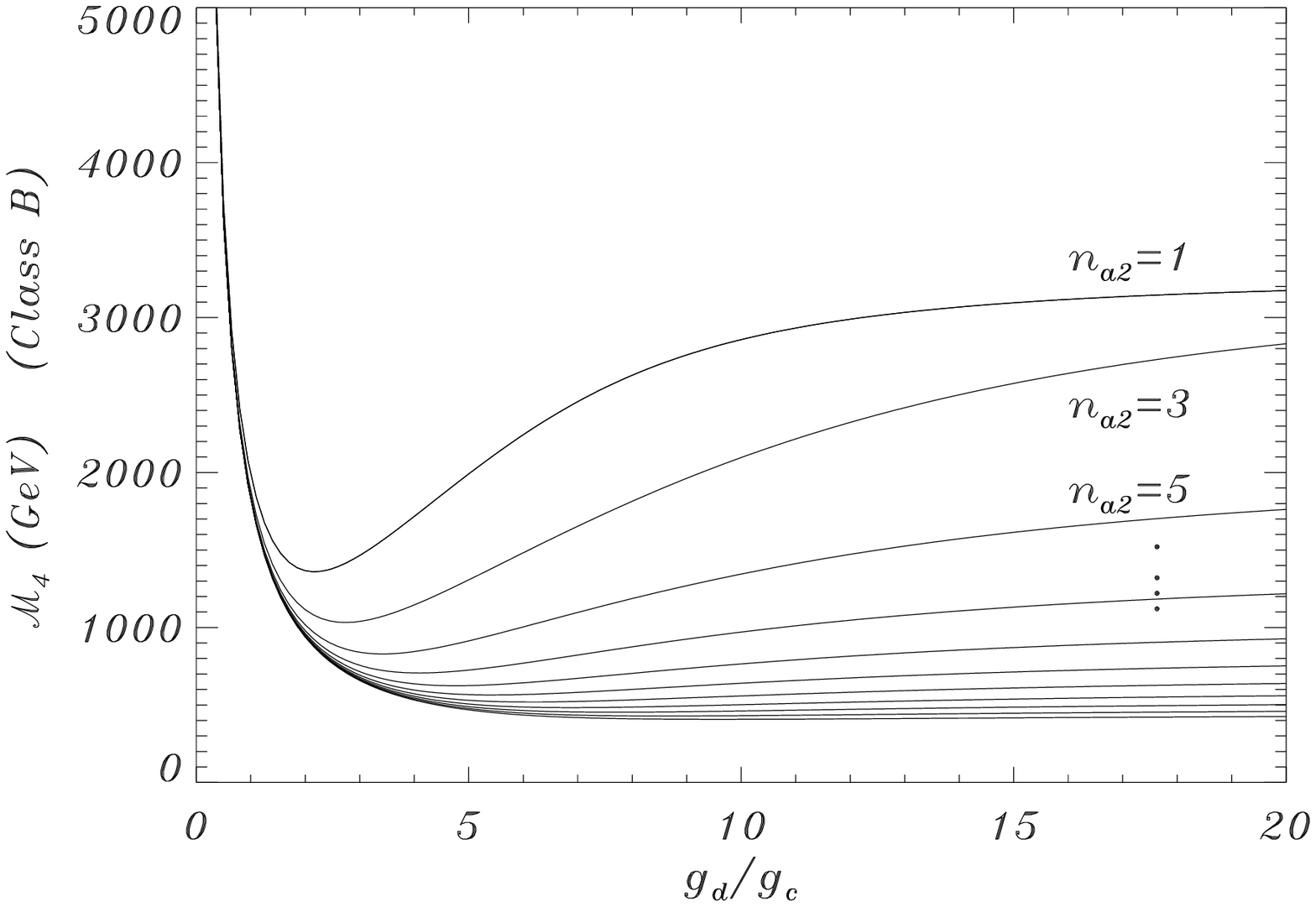,height=5.4cm,width=5.66cm}}  
\end{tabular} 
\end{center}
\caption{\small{\newline
{\bf Left column:} Masses of $U(1)$ fields (GeV) in Class A models 
 with $\beta_1=1/2$, $\beta_2=1/2$, $\nu=1/3$
$n_{c1}=1$, $n_{b1}=-1$ in function  of the parameter
ratio $g_d/g_c$ for varying wrapping number $n_{a2}$.
The lower bound on $\cM_4$  is saturated at large $n_{a2}$.\newline
{\bf Right column:} Masses of $U(1)$ fields (GeV) in Class B models 
with $\beta_1=1$, $\beta_2=1$, $\nu=1$, $n_{c1}=1$, $\phi=\pi/6$.
in function of $g_d/g_c$ for varying wrapping number  $n_{a2}$.
The lower bound on $\cM_4$ is saturated at large $n_{a2}$, but
corresponds to a mixing angle beyond $1.5\times 10^{-3}$. 
$U(1)_b$ of mass $\cM_3$ does not mix with the rest of 
$U(1)$ fields (before electroweak
symmetry breaking).  These mass values comply with  
$\rho$ parameter constraints and include their electroweak corrections.}}
\label{f1}
\end{figure} 
Even in the case of more than one additional boson one may in
principle define an {\it effective} angle $\psi^{'}_i$ for the 
mixing of $Z$ boson with a massive
$A_i'$ state (i=2,3,4)  as in (\ref{oneboson}) with $M_{Z'}\ra \cM_i$,
$i=2,3,4.$ This  may further be expressed as 
\begin{equation}\label{teff}
\psi'_i= \arctan
\left[\frac{\Delta\rho/\rho_0}{\Delta\rho/\rho_0+1}\frac{1}
{\cM_i^2/M_0^2-1}\right]^{1/2},\qquad i=2,3,4.
\end{equation}
where we used that for the $\rho$ parameter \cite{pdg}
$\rho_0=M_W^2/(M_0^2 \cos\theta_W)$ and thus
$\Delta\rho/\rho_0=-1+M_{0}^2/M_Z^2$.
Detailed comparison of this amount of mixing with that of (\ref{zz'})
shows that $\psi_i^{'}$  may provide an estimate for the 
amount of mixing for the cases we considered, but it is generically 
larger than that of (\ref{zz'}). 
Also definition (\ref{teff}) does not contain information about the
sign of the mixing  and its dependence on the parameters of the
model ($n_{a2}$ and $g_d/g_c$) is different from that of 
eq.(\ref{zz'}) which will be used throughout this analysis.

From the mass eigenvalue equation after electroweak symmetry breaking
 $det(\cM^2-\cM_i^2 I_5)=0$, $i={\overline{1,5}}$ one finds $\cM_i^2$
in terms of the string scale, $M_S$. Using  the lower bounds 
(\ref{msnew}) on the latter we can then make predictions for 
(lower bounds on) the masses of
additional $U(1)$ fields $A_i^*$ (i=2,3,4). These bounds are therefore 
{\it independent} of the volume factors ${\underline \xi}^i$, entering in 
the ``re-scaled'' value of $M_S$.  The results are presented
in Table \ref{tablespectrum} and discussed below in function of
the  associated mixing.

\begin{figure}[t] 
\begin{center}
\begin{tabular}{cc} 
\parbox{7.1cm}{\psfig{figure=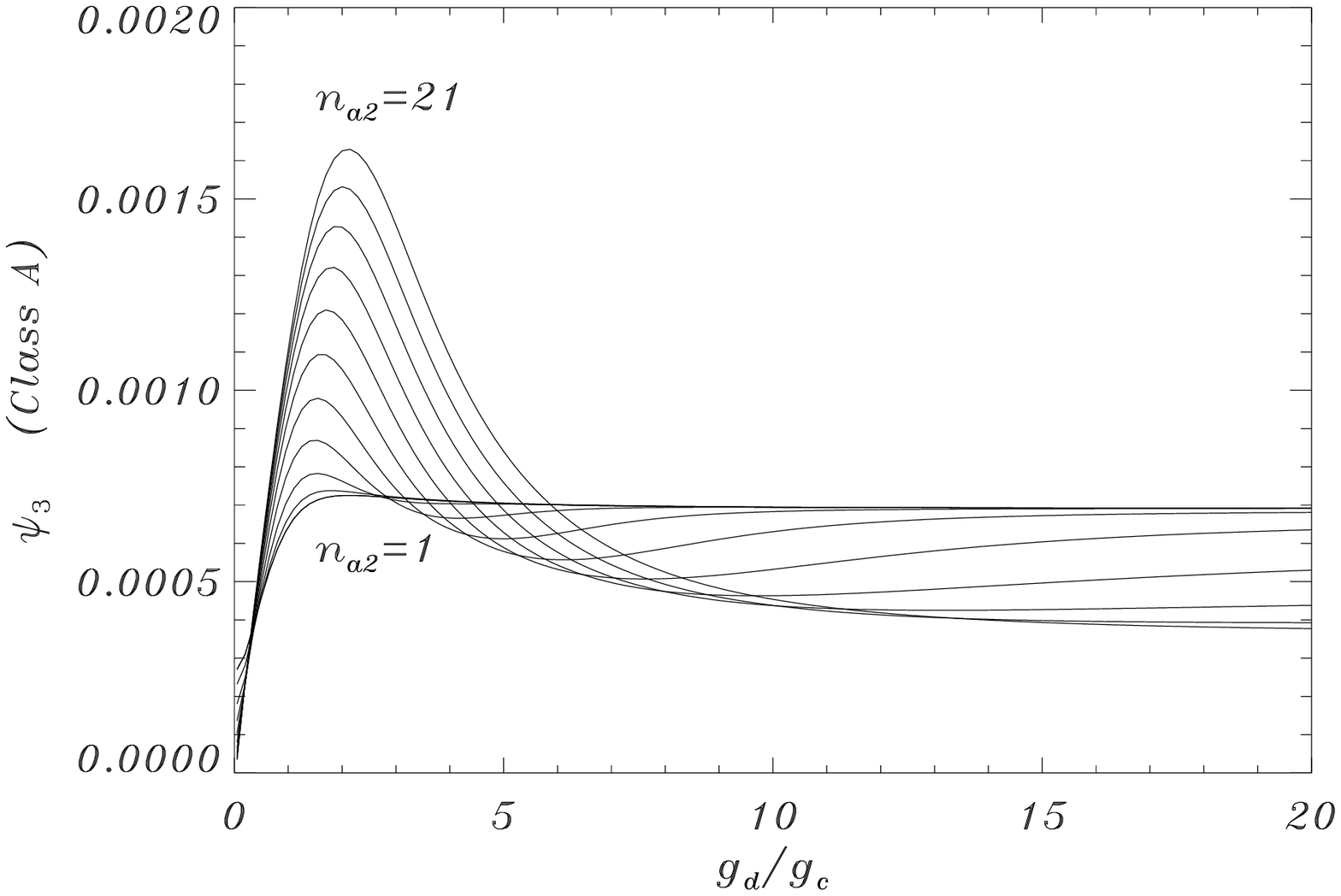,height=6cm,width=6.3cm}}
\end{tabular} 
\end{center}
\caption{\small{
Class A models:  
($\beta_1=1/2$, $\beta_2=1/2$, $\nu=1/3$ $n_{c1}=1$, $n_{b1}=-1$). 
The mixing $\psi_3$ of Z boson with $A_3'$ the lightest state among the 
additional $U(1)$'s. $n_{a2}$ varies as shown with step 2.
The largest amount of mixing is usually manifest for $g_d\approx g_c$, 
increasing  with $n_{a2}$. Upper bounds on the former translate into 
bounds on $n_{a2}$ and thus on the $U(1)$ masses, Figure \ref{f1}.
}}\label{f2}
\end{figure}

\begin{table}[ht] 
\begin{center}
\begin{tabular}{|c|c|c|c|c|c|c|c|c|}
\hline
Higgs        & $\nu$ & $\beta_1$ & $\beta_2$ & $n_{c1}$ & $n_{b1}$ & $\cM_2\, (TeV)$  & $\cM_3\, (TeV)$  & $\cM_4\,(TeV)$  \\
\hline\hline  
$n_H=1,n_h=0$   & $1/3$   &  1/2  &   $1/2$  & $1$     &  $-1$   & $>25$  & $1.2 \, (0.75)$ &  $>3.5$  \\
$n_H=1,n_h=0$   & $1/3$   &  1/2  &   $1$    & $1$     &  $-1$   & $>110$ & $1.2\,(0.65)$   &  $>3.5$  \\
$n_H=1,n_h=0$   & $1/3$   &  1/2  &   $1/2$  & $-1$    &  $1$    & $>25$  & $1.7\,(0.75)$   &  $>3.5$  \\
$n_H=1,n_h=0$   & $1/3$   &  1/2  &   $1$    & $-1$    &  $1$    & $>100$ & $1.2\,(0.65)$   &  $>3.5$  \\
$n_H=0,n_h=1$   & $1/3$   &  1/2  &   $1/2$  & $1$     &  $1$    & $>25$  & $1.2\,(0.6)$    &  $>3.5$  \\
$n_H=0,n_h=1$   & $1/3$   &  1/2  &   $1$    & $1$     &  $1$    & $>100$ & $1.2\,(0.6)$    &  $>3.5$  \\
$n_H=0,n_h=1$   & $1/3$   &  1/2  &   $1/2$  & $-1$    &  $-1$   & $>25$  & $1.1\,(0.65)$    &  $>3.5$  \\
$n_H=0,n_h=1$   & $1/3$   &  1/2  &   $1$    & $-1$    &  $-1$   & $>65$  & $1.2\,(0.6)$    &  $>3.25$  \\

\hline\hline  
$n_H=1,n_h=1$   & $1$      &  1  &   $1/2$   & $1$    &  $0$   & $>1.5$   & $>1.5$ & $1\,(0.6)$     \\
$n_H=1,n_h=1$   & $1$      &  1  &   $1$     &  $1$   &  $0$   & $>4.5$   & $>1.5$ & $1.2\,(0.6)$   \\
$n_H=1,n_h=1$   & $1$      &  1  &   $1/2$   &  $-1$  &  $0$   & $>3$     & $>1.7$ & $0.75\,(0.6)$   \\
$n_H=1,n_h=1$   & $1$      &  1  &   $1$     &  $-1$  &  $0$   &  $>5.4$  & $>1.6$ & $1.1\,(0.55)$    \\
$n_H=1,n_h=1$   & $1/3$    &  1  &   $1/2$   &  $1$   &  $0$   &  $>3.5$  & $>1.6$ & $1.2\,(0.6)$    \\
$n_H=1,n_h=1$   & $1/3$    &  1  &   $1$     &  $1$   &  $0$   &  $>12.5$ & $>1.6$ & $1.125\,(0.6)$   \\
$n_H=1,n_h=1$   & $1/3$    &  1  &   $1/2$   &  $-1$  &  $0$   &  $>3.6$  & $>1.6$ & $1.1\,(0.575)$   \\
$n_H=1,n_h=1$   & $1/3$    &  1  &   $1$     &  $-1$  &  $0$   &  $>13$   & $>1.7$ & $1.125\,(0.6)$   \\
\hline 
\end{tabular}
\end{center}
\caption{\small Lower bounds for Class A (Class B) models in the upper
(lower) table. The values correspond to any of Z bosons' mixing with
massive U(1)'s  less than $1.5\times 10^{-3}$. Values in brackets
correspond to mixings of up to $3\times 10^{-3}$.(For Class B models
we chose  $\theta=\pi/6$).
 These bounds are only reached in cases with $g_d/g_c=\cO(1)$, 
otherwise they may increase further.}
\label{tablespectrum}
\end{table}

Model independent constraints on the value of the mixing lead to 
$|\psi|<0.003$ \cite{Leike:1998wr}, but there are cases when this may 
be  smaller ($10^{-3}$) \cite{Erler:1999nx}. We thus presented in
Table \ref{tablespectrum} our
results for two (upper) values of $1.5\times 10^{-3}$ and $3\times
10^{-3}$ of the mixing angles of Z boson with {\it any} massive $U(1)$ field.
In Figure \ref{f1} lower bounds on the masses of $U(1)$ fields are
also presented. 
For fixed $g_d/g_c$ the bounds on the mixing angle set bounds on
$n_{a2}$ which is then used for finding the lowest allowed values
for the $U(1)$ masses. 
For Class A models, Figure \ref{f1} (left column) one finds that
generically $\cM_2>25\times 10^3$ GeV and $\cM_4>3.5$ TeV. $\cM_3$ can
be as low as 1200 GeV for  a mixing  of order 
$\psi_3\approx 1.5 \times 10^{-3}$ with $g_d/g_c=\cO(1)$, 
see also Figure \ref{f2}. For $\psi_3\approx 3 \times 10^{-3}$
the bound on $\cM_3$ decreases to  $\cM_3>750$ GeV. For Class A models 
with $\beta_2=1$ instead  of $\beta_2=1/2$ the lowest
bounds on $\cM_3$ and $\cM_4$ do not change significantly, while for $\cM_2$
an  increase of factor $\approx 4$ is present while still keeping the
same amount of mixing.
Changing the sign of $n_{c1}$ and $n_{b1}$ for fixed $\beta_{1,2}$ and
$\nu$ does not affect significantly these bounds.

For Class B models, Figure \ref{f1} (right) the lower bounds on the
masses have a similar dependence in function of $g_d/g_c$ giving
$\cM_2>4.5$ TeV, $\cM_3>1.5$ TeV and $\cM_4>1200$ GeV
corresponding to a mixing of $1.5 \times 10^{-3}$ (or to 
$n_{a2}=2$).  $\cM_4$  decreases to 600 GeV for $\psi=3 \times 10^{-3}$.
$\cM_3$ is proportional to $M_S$ and 
is the analogue of the anomalous $U(1)$ in heterotic
models (it does not mix with the remaining U(1)'s eq.(\ref{eigvecB}), 
unlike the case of Class A models \cite{Ghilencea:2002da}).
Changing $\beta_2$, $n_{c1}$, $n_{b1}$ and $\rho$ brings in
(small) changes on the bounds, as presented in Table 
\ref{tablespectrum}.

To conclude, the lowest bounds for $U(1)$ masses with any of their 
associated mixing angles with Z boson
in the region of $1.5 \times 10^{-3}$ or less, 
are  1100 GeV for Class A models and 750 GeV for Class B models.
Note  that for a fixed amount of mixing, when the two $U(1)$ 
couplings $g_c,g_d$ are comparable,  
one finds for generic cases lower bounds on masses than when
the couplings are  significantly different.
These bounds should be compared to current   
experimental $Z'$ mass limit which is $>690$ GeV and
was obtained by CDF with the assumption that the $Z'$ boson has SM 
couplings strengths \cite{Abe:1997fd}. 
To help identify  specific signatures of new $U(1)$ fields and  
distinguish from other models with additional $Z'$ bosons, 
the full Z boson eigenvector is presented in the Appendix. 
The mass bounds on the additional $U(1)$
bosons we found are somewhat larger than those of alternative 
models for similar amounts of mixing, which are  
\cite{pdg} in the range of 545 GeV (SO(10)
GUT models), 564 GeV (for left-right models with gauge group $SU(3)_c\times
SU(2)_L\times SU(2)_R\times U(1)_{B-L}\subset SO(10)$ and 809 GeV
for the sequential $Z_{SM}$ boson defined to have the same 
couplings to fermions as the SM Z boson.

\section{Conclusions}
The analysis of phenomenological viability of consistent D-brane 
SM-like models is at an early stage, 
and this work was intended as a step in this direction. 
In such models,  additional massive $U(1)$ fields are a generic presence.
Previous analysis of the implications of these  $U(1)$'s and of 
the value of the string scale  revealed that $\rho$ parameter constraints 
can be respected  for a   string scale  in the TeV region. 
This result was  used in this work to set lower bounds
on the masses of the additional $U(1)$'s, independent of the
volume factors $\underline\xi^i$ affecting the string scale prediction.
The  masses of the $U(1)$ fields are of string origin (with small electroweak 
corrections suppressed by the string scale), therefore
no additional Higgs states (beyond the  SM case) are  required.
The values of $U(1)$ masses were found to be somewhat larger than those of 
alternative models.  The  amount of  mixing of the SM Z boson with any of the
new $U(1)$ fields  and the eigenvectors of the  $U(1)$ fields were
computed.  This information may be further used to 
improve our  bounds on  the $U(1)$ masses from (upper) bounds on the mixing 
derived from the dilepton decay modes of $Z'$ bosons.

\vspace{0.6cm}
\noindent
{\bf Acknowledgements:} 
The author thanks  L.~E.~Ib\'a\~nez and  F. Quevedo 
for helpful discussions and comments on this work. 
He  thanks N. Irges and  R.~Rabad\'an 
for many discussions on the model \cite{imr}.
This work was supported by PPARC (U.K.).

\newpage
\def\theequation{A-\arabic{equation}}
\appendix
\section*{Appendix:}

{\bf $D6$ brane models:} The $U(1)_\alpha$ charges  of quarks and leptons 
in D6-brane models of \cite{imr}  are:
\begin{table}[ht]
\begin{center}
\begin{tabular}{|c|c|c|c|c|c|c|c|}
\hline Intersection &
 Matter fields  &   &  $q_a$  & $q_b $ & $q_c $ & $q_d$  & $q_Y$ \\
\hline\hline (ab) & $Q_L$ &  $(3,2)$ & 1  & -1 & 0 & 0 & 1/6 \\
\hline (ab*) & $q_L$   &  $2( 3,2)$ &  1  & 1  & 0  & 0  & 1/6 \\
\hline (ac) & $U_R$   &  $3( {\bar 3},1)$ &  -1  & 0  & 1  & 0 & -2/3 \\
\hline (ac*) & $D_R$   &  $3( {\bar 3},1)$ &  -1  & 0  & -1  & 0 & 1/3 \\
\hline (bd*) & $ L$    &  $3(1,2)$ &  0   & -1   & 0  & -1 & -1/2  \\
\hline (cd) & $E_R$   &  $3(1,1)$ &  0  & 0  & -1  & 1  & 1   \\
\hline (cd*) & $N_R$   &  $3(1,1)$ &  0  & 0  & 1  & 1  & 0 \\
\hline 
\end{tabular}
\end{center}
 \caption{\small  The hypercharge generator is defined  by
 $q_Y = 1/6 \,q_a - 1/2 \,q_c +  1/2\, q_d$.
The asterisk denotes the ``orientifold mirror'' of each given brane. 
$U(1)_a$ and $U(1)_d$ can be identified with 
baryon number and (minus) lepton number respectively. 
$U(1)_c$ can be identified with the third component 
of right-handed weak isospin. $U(1)_b$ is an axial symmetry
with QCD anomalies, much like a PQ-symmetry.
$U(1)_b$ and $3U(1)_a-U(1)_d$ linear combination 
have triangle anomalies, cancelled by a generalised Green-Schwarz
mechanism, whereas $U(1)_a+3U(1)_d$ and $U(1)_c$ are anomaly-free.}
\label{tabpssm}
\end{table}

\vspace{0.3cm}
\noindent
The coefficients $c_i^\alpha$ 
encountered in the text eq.(\ref{masones}) are given by
\cite{imr}
\begin{equation}
c_i^\alpha \ =\ N_\alpha  n_{\alpha j} \, n_{\alpha k} \, m_{\alpha i} 
\ ;\  i\not= j\not= k\not= i  \ , \ i=1,2,3
\label{cillos}
\end{equation}
$N_\alpha$ is the number of parallel branes of type $\alpha $. 
The  wrapping numbers $n_\alpha$, $m_\alpha$ 
as derived in \cite{imr} are:

\begin{table}[htb] \footnotesize
\begin{center}
\begin{tabular}{|c||c|c|c|}
\hline
 $N_\alpha$    &  $(n_{\alpha 1},m_{\alpha 1})$  &
 $(n_{\alpha 2},m_{\alpha 2})$   & $(n_{\alpha 3},m_{\alpha 3})$ \\
\hline\hline $N_a=3$ & $(1/\beta _1,0)$  &  $(n_{a 2},\epsilon \beta_2)$ &
 $(1/\nu ,  1/2)$  \\
\hline $N_b=2$ &   $(n_{b 1},-\epsilon \beta_1)$    &  $ (1/\beta_2,0)$  &
$(1,3\nu /2)$   \\
\hline $N_c=1$ & $(n_{c 1},3\nu \epsilon \beta_1)$  &
 $(1/\beta_2,0)$  & $(0,1)$  \\
\hline $N_d=1$ &   $(1/\beta_1,0)$    &  $(n_{d 2},-\beta_2\epsilon/\nu )$  &
$(1, 3\nu /2)$   \\
\hline \end{tabular}
\end{center}
 \caption{\small{ D6-brane wrapping numbers giving rise to a SM
spectrum \cite{imr}.
The general solutions yielding the SM spectrum
 are parametrized by a phase $\epsilon =\pm1$, the NS background
on the first two tori $\beta_i=1-b_i=1,1/2$, four integers
$n_{a 2},n_{b 1},n_{c 1},n_{d 2}$ and a parameter $\nu=1,1/3$.}
\label{solution} }
\end{table}
\vspace{0.4cm}
\noindent
{\bf Eigenvectors in Class A models.}
For Class A models the matrix $\cF_{i\alpha}$ ($i=1,2,3,4$, $\alpha=a,b,c,d$) is
given below (with $\lambda_1=0$ (hypercharge) and
$\lambda_i=M_i^2/M_S^2$, $i=2,3,4$ the roots of $\det(\lambda M_S^2
I_4-M^2)=0$, computed in \cite{Ghilencea:2002da})
\begin{eqnarray}\label{eigvecA}
\cF_{1\alpha}&=&\frac{1}{|\cF_1|} \left\{\frac{g_d}{3 g_a}, 0,
-\frac{g_d}{g_c}, 1\right\}_\alpha,\qquad {\alpha=a,b,c,d} \nonumber \\ 
\nonumber\\
\cF_{i a}&=& \frac{1}{|\cF_i|}
\frac{3 \beta_2\, g_a \left( 2 \beta_2 \,g_d^2\, n_{c1}  - 
 n_{a2} \gl_i \beta_1\nu^2\right)}
{g_d\left[ 18 \beta_2^2 \, g_a^2 n_{c1}  + \beta_1 \nu^2 \gl_i (n_{a2}
\beta_2 -2 \beta_1 n_{c1})\right]},\qquad\qquad i=2,3,4. \nonumber \\ 
\nonumber\\
\cF_{ib}&=&\frac{1}{|\cF_i|} \left\{- \frac{4 \beta_1^2\,  \beta_2^2 \, \nu^2 \gl_i^2 
+4 \beta_2^2  \left[g_c^2 g_d^2 +9 g_a^2 (g_c^2+g_d^2)\right]n_{c1}^2 }
{6 \nu\, n_{b1}\, g_b\, g_d\left[ 18 \beta_2^2\, g_a^2 n_{c1} +\beta_1
\nu^2 \gl_i (n_{a2} \beta_2 -2 \beta_1 n_{c1})\right] }\right.\nonumber\\   
 \nonumber\\
&&\left.+\frac{\gl_i\left[\beta_2^2 (9 g_a^2+g_d^2)(4  \beta_2^2  +\nu^2 n_{a2}^2)
-4 \beta_1 \nu^2 n_{c1} (\beta_2  g_d^2 n_{a2} - \beta_1 n_{c1}
(g_c^2+g_d^2) )\right]}
{6 \nu\, n_{b1}\, g_b\, g_d\left[ 18 \beta_2^2\,  g_a^2\,  n_{c1} +\beta_1
\nu^2 \gl_i (n_{a2} \beta_2 -2 \beta_1 n_{c1})\right]} \right\}, \nonumber \\
\nonumber\\
\cF_{ic}&=&\frac{1}{|\cF_i|} 
\frac{2 g_c n_{c1} \left[\beta_2^2 \, \, (9 g_a^2+g_d^2) 
-\beta_1^2\nu^2 \gl_i\right]}
{g_d\left[ 18 \beta_2^2 g_a^2 n_{c1}   +\beta_1 \nu^2 \gl_i (n_{a2}
\beta_2 -2 \beta_1 n_{c1})\right]},\qquad\qquad
\cF_{id}=\frac{1}{|\cF_{i}|}
\end{eqnarray}
with the notation $|\cF_i|\equiv (\sum_{\alpha} \cF_{i\alpha}^2)^{1/2}$.

\vspace{1cm}
\noindent
{\bf Eigenvectors in Class B models.} For 
Class B models the matrix $\cF_{i\alpha}$ is given by:
\begin{eqnarray}\label{eigvecB}
\cF_{1\alpha}&=&\frac{1}{|\cF_1|} \left\{\frac{g_d}{3 g_a}, 0,
-\frac{g_d}{g_c}, 1\right\}_\alpha,\qquad {\alpha=a,b,c,d} \nonumber \\
\cF_{2\alpha}&=&\frac{1}{|\cF_2|}\left\{\frac{3 g_a}{g_d}\frac{1}{2
\omega_1}\left(\omega_2-8 \beta_1^2\beta_2^2\nu^2 y\right), 0,
\frac{g_c}{g_d}\frac{1}{2\omega_1}\left(\omega_3- 8
\beta_1^2\beta_2^2\nu^2 y\right),1\right\}_\alpha,\nonumber \\
\cF_{3\alpha}&=&\left\{0,1,0,0\right\}_\alpha, \nonumber \\
\cF_{4\alpha}&=&\frac{1}{|\cF_4|}\left\{\frac{3 g_a}{g_d}\frac{1}{2
\omega_1}\left(\omega_2+8 \beta_1^2\beta_2^2\nu^2 y\right), 0,
\frac{g_c}{g_d}\frac{1}{2\omega_1}\left(\omega_3+ 8
\beta_1^2\beta_2^2\nu^2 y\right),1\right\}_\alpha
\end{eqnarray}
where 
\begin{eqnarray}
\omega_1&=&36 \beta_2^4  g_a^2+\nu^2\left(\beta_2n_{a2}-2 \beta_1
n_{c1}\right)\left(9 \beta_2 g_a^2 n_{a2}+2 \beta_1 n_{c1} g_c^2\right),
 \nonumber \\
&&\vspace{0.3cm}\hfill\nonumber\\
\omega_2&=&-\left\{\beta_2^2(9 g_a^2-g_d^2)(4 \beta_2^2+\nu^2
n_{a2}^2)+4 \nu^2 \beta_1n_{c1}(g_c^2+g_d^2)(n_{a2} \beta_2-\beta_1
n_{c1})\right\}, \nonumber \\
&&\vspace{0.3cm}\hfill\nonumber \\
\omega_3&=&\left\{\beta_2(9 g_a^2+g_d^2)(4 \beta_2^3+\nu^2\beta_2
n_{a2}^2-4 \nu^2\beta_1 n_{a2} n_{c1})
+4\nu^2\beta_1^2(g_d^2-g_c^2) n_{c1}^2\right\}\nonumber
\end{eqnarray}
with the notation:
\begin{eqnarray}\label{spect}
y&=&\left\{x^2-\frac{ n_{c1}^2}{\beta_1^2\nu^2}
\left[g_c^2 g_d^2+9 g_a^2(g_c^2+g_d^2)\right]\right\}^{1/2}\nonumber \\
x&=&\frac{1}{8\beta_1^2 \beta_2^2\nu^2}\left\{
\left[9g_a^2+g_d^2\right]\left[4
\beta_2^4+\nu^2\beta_2^2n_{a2}^2\right]+4\beta_1
n_{c1}\nu^2 \left[\beta_1(g_c^2+g_d^2) n_{c1}-\beta_2
g_d^2 n_{a2}\right]\right\}
\end{eqnarray}
The  eigenvectors $\cF_{i\alpha}$ (i=fixed), 
correspond to mass eigenvalues $M_i^2$ with $M_1^2=0$, $M_2^2=(x+y)
M_S^2$, $M_3^2=(2\beta_1/\beta_2)^2 g_b^2 M_S^2$ and $M_4^2=(x-y)
M_S^2$ respectively.

\vspace{1cm}
\noindent
{\bf Eigenvectors after Electroweak Symmetry Breaking.}
The matrix $\cF_{i\gamma}^*$ (with $i={\overline{1,5}}$ and $\gamma=a,b,c,d,W_3$) 
after electroweak symmetry breaking has the same structure 
for both Class A and Class B models and is given below. 
$\lambda_i^*=\cM_i^2/M_S^2$ is an eigenvalue of $\cM^2_{\gamma\gamma'}$,
expressed in string units, $\lambda_1^*=0$ (photon state) 
and  $\delta\equiv \cos(2\theta)$ with
$\theta$ defined in the text as the mixing angle in the Higgs sector. 
For Class B models one should set  in the eqs. below $n_{b1}=0$. 
\begin{eqnarray}\label{EWSB}
\cF_{1\gamma}^*&\!\!\!\!\!\!\!\!=& \!\!\!\!\!\!
\frac{1}{|\cF_{1}^*|}\left\{\frac{g_d}{3 g_a}, 
0, -\frac{g_d}{g_c},1,-\frac{g_d}{g_b}\right\}_{\gamma}, \qquad
\gamma=a,b,c,d,W_3.\\
\nonumber\\
\cF_{ia}^*&\!\!\!\!\!\!\!\!=&\!\!\!\!\!\! \frac{1}{|\cF_i^*|}
\frac{3 \beta_2\, g_a \left( 2 \beta_2 \,g_d^2\, n_{c1}  - 
 n_{a2} \gl_i^* \beta_1\nu^2\right)}{
g_d\left[ 18 \beta_2^2 \, g_a^2 n_{c1}  + \beta_1 \nu^2 \gl_i^* (n_{a2}
\beta_2 -2 \beta_1 n_{c1})\right]},\qquad \qquad  i=2,3,4,5.\nonumber \\ 
\nonumber\\
\cF_{ib}^*&\!\!\!\!\!\!\!\! =& \!\!\! \!\!\! \!\! \left\{
( \eta (g_b^2+g_c^2)-\lambda_i^*)\! 
\left[4 \beta_2^4 
(9 g_a^2 +\! g_d^2) \lambda_i^* -\! 4 \beta_1 \beta_2 g_d^2\lambda_i^*
n_{a2} n_{c1} \nu^2+ \nu^2\lambda_i^* \beta_2^2
((9g_a^2+g_d^2)n_{a2}^2- \! 4 \beta_1^2\lambda_i^* )
\right]\right.\nonumber\\
&+&\!\!\!\! \beta_2^2\left[4 n_{c1}^2 (g_c^2 g_d^2+9 g_a^2 g_d^2)(\lambda_i^*-g_b^2
\eta) + 9 g_a^2 g_c^2 (\lambda_i^*-(g_b^2+g_d^2)\eta)\right]\nonumber\\
&+&\!\!\!\! \left.  4 \beta_1^2 \lambda_i^* n_{c1}^2 \nu^2 
\left[(-\lambda_i^*+g_b^2 \eta) (g_c^2+g_d^2)+g_c^2 g_d^2
\eta\right]\right\}\!
\left\{ 2 g_b g_d \left[18 \beta_2^2 g_a^2 n_{c1}\right.  \right.\nonumber\\
&-& \!\!\! \! \left.\left. \beta_1 \lambda_i^*
\left(-\beta_2 n_{a2}+2 \beta_1 n_{c1}\right)\nu^2\right]
\times \left[-3 \lambda_i^* n_{b1}\nu +3 g_b^2 n_{b1} \eta \nu +g_c^2 \eta (3
n_{b1} \nu -n_{c1} \delta)\right]\right\}^{-1}
|\cF_{i}^*|^{-1} \nonumber \\
\nonumber \\
\cF_{ic}^*&\!\!\!\!\!\!\!\!=&\!\!\! \!\!\!\! -g_c\left[4 \beta_2^4 
(9 g_a^2 +g_d^2)\lambda_i^* \eta
\delta-4 \beta_1 \beta_2 g_d^2\lambda_i^* n_{a2} n_{c1} \eta \nu^2 \delta+
4 \beta_1^2 \lambda_i^* n_{c1}\nu^2 
(3 g_b^2 n_{b1} \eta \nu+ g_d^2 n_{c1} \eta \delta \right.  \nonumber\\
&-&\left. 3 \lambda_i^* n_{b1} \nu) +
 \beta_2^2 \left[-4 \beta_1^2 \lambda_{i}^{* 2} \eta \nu^2 \delta - 12 n_{c1} \eta
\left(g_b^2 n_{b1} \nu (9 g_a^2 +g_d^2) +3 g_a^2 g_d^2 n_{c1} \delta
\right) \right. \right. \nonumber\\
&+&\!\!\!\! 
\left.\left. (9 g_a^2 +g_d^2) \lambda_i^* \nu (12 n_{b1} n_{c1} +n_{a2}^2 \eta \nu
\delta )\right]\right]
\left\{2 g_d \left[18 \beta_2^2 g_a^2 n_{c1}+\beta_1\lambda_i^* \nu^2 (\beta_2
n_{a2}-2 \beta_1 n_{c1})\right]\right.\nonumber\\
&\times &\!\!\!\! \left.  \left[-3 \lambda_i^* n_{b1} \nu +3 g_b^2 n_{b1} \eta \nu + g_c^2 \eta (3
n_{b1} \nu -n_{c1} \delta)\right]\right\}^{-1}
|\cF_{i}^*|^{-1}\nonumber\\
\cF_{id}^*&\!\!\!\!\!\!\!\!=&\!\! |\cF_i^*|^{-1}\nonumber\\
\cF_{i {W_3}}^*&\!\!\!\!\!\!\!=&\!\!\!\!\! \! 
g_b \eta \left[12 g_c^2 n_{b1} n_{c1} \nu  \left(\beta_2^2 (9 g_a^2\! +
\! g_d^2)-\beta_1^2 \lambda_i^* \nu^2\right)+
\! \left[4 \beta_2^4 (9 g_a^2\!
 + \! g_d^2) \lambda_i^* -4 \beta_2^2 n_{c1}^2 (g_c^2 g_d^2 +9
g_a^2 (g_c^2 +g_d^2)) \right. \right.\nonumber\\
&-&\!\!\!\! \left.\left. \lambda_i^* \left[4 \beta_1 \beta_2 g_d^2 n_{a2}
n_{c1} - \beta_2^2 n_{a2}^2 (9 g_a^2 +g_d^2) + 4 \beta_1^2
\left(\beta_2^2 \lambda_i^* -(g_c^2 + g_d^2) n_{c1}^2
\right)\right]\nu^2\right]\delta\right]|\cF_{i}^*|^{-1}\nonumber\\
&\times& \! \!\!\!\!\! \left\{2 g_d \left[18 \beta_2^2 g_a^2
n_{c1}+\! \beta_1\lambda_i^* \nu^2 (\beta_2 n_{a2}-2 \beta_1 n_{c1})\right]
\left[-3 \lambda_i^* n_{b1} \nu +3 g_b^2 n_{b1} \eta \nu + g_c^2 \eta (3
n_{b1} \nu\! -\! n_{c1} \delta)\right]\right\}^{-1}\nonumber
\end{eqnarray}
where we used the notation $|\cF_i^*|^2=
\sum_{\gamma: a,b,c,d,W_3} |\cF_{i\gamma}^*|^2$
and where one has that \cite{Ghilencea:2002da}
\begin{equation}\label{hypercharge1}
\frac{1}{g_y^{2}}=\frac{1}{36}\frac{1}{g_a^2}+
\frac{1}{4}\frac{1}{g_c^2}+\frac{1}{4}\frac{1}{g_d^2}
\end{equation}
with $g_y$ the hypercharge coupling. Eq.(\ref{hypercharge1})
with fixed $g_y$ and $g_a^2=g^2_{QCD}/6$ establishes 
a correlation for the allowed values of $g_c$ and $g_d$
and this is used in the text.
To compute the mixing of $Z$ boson with the massive $A'_i$ fields
($i=2,3,4$) one replaces $\lambda_i^*$ with $\cM_5^2/M_S^2$ where $\cM_5^2$
and $M_S^2$ were given in the text.

\end{document}